\documentclass[12pt]{article}

\input epsf.sty
\usepackage{amsmath,amssymb}
\usepackage{amsfonts}
\newcommand{\be}{\begin{equation}}
\newcommand{\ee}{\end{equation}}
\newcommand{\bea}{\begin{eqnarray}}
\newcommand{\eea}{\end{eqnarray}}
\newcommand{\nn}{\nonumber}

\def\CO{{\cal O}}

\newcommand{\bra}[1]{\langle{#1}|}
\newcommand{\ket}[1]{|{#1}\rangle}

\newcommand{\tr}[1]{{\rm tr}[{#1}]}

\newcommand{\AdS}[1]{{\rm AdS}_{#1}}

\newcommand{\myfig}[3]{\begin{figure}[ht]
\begin{center}
\leavevmode
\epsfxsize=#2cm
\epsfbox{#1}
\end{center}
\caption{#3}
\label{fig:#1}
\end{figure}}

\begin{document}

\rightline{hep-th/0505123}
\rightline{DCTP-05/21}
\rightline{UPR-1124-T}

\centerline{\Large \bf }\vskip0.25cm
\centerline{\Large \bf }\vskip0.25cm
\centerline{\LARGE \bf The Library of Babel}\vskip0.25cm

\vskip 1cm
\renewcommand{\thefootnote}{\fnsymbol{footnote}}
\centerline{{\bf 
Vijay Balasubramanian,${}^{1}$\footnote{vijay@physics.upenn.edu}
Vishnu Jejjala,${}^{2}$\footnote{vishnu.jejjala@durham.ac.uk}
and
Joan Sim\'on${}^{1}$\footnote{jsimon@bokchoy.hep.upenn.edu}
}}
\vskip .5cm
\centerline{${}^1$\it David Rittenhouse Laboratories}
\centerline{\it University of Pennsylvania}
\centerline{\it Philadelphia, PA 19104, U.S.A.}
\vskip .5cm
\centerline{${}^2$\it Centre for Particle Theory}
\centerline{\it Department of Mathematical Sciences}
\centerline{\it University of Durham}
\centerline{\it South Road, Durham DH1 3LE, U.K.}
\vskip .5cm

\vskip 1cm
\begin{abstract}
We show that heavy pure states of gravity can appear to be mixed states to almost all probes.
Our arguments are made for $\rm{AdS}_5$ Schwarzschild black holes using the field theory dual to string theory in such spacetimes.
Our results follow from applying information theoretic notions to field theory operators capable of describing very heavy states in gravity.
For certain supersymmetric states of the theory, our account is exact:
the microstates are described in gravity by a spacetime ``foam'', the precise details of which are invisible to almost all probes.
\end{abstract}

\setcounter{footnote}{0}
\renewcommand{\thefootnote}{\arabic{footnote}}

\newpage

{\small
\begin{quotation}
{\em There are twenty-five orthographic symbols. That discovery enabled
mankind, three hundred years ago, to formulate a general theory of the
Library and thereby satisfactorily solve the riddle that no conjecture
had been able to divine --- the formless and chaotic nature of virtually
all books....
The Library is total and its shelves register all the possible combinations of the twenty-odd
orthographic symbols (a number which, though extremely vast, is not infinite).} \\
%{\em Infidels claim that the rule in the Library is not ``sense,'' but ``non-sense,'' 
%and that ``rationality'' (even humble, pure coherence) is an almost miraculous exception.} \\
\rightline{--- Jorge Luis Borges, in ``The Library of Babel'' $\;\;\;\;\;\;\;\;\;\;\;\;\;\;\;\;\;\;\;\;\;\;$}
\end{quotation}}

${}$

\noindent
The microcanonical accounting of the entropy of black holes posits an enormous number of degenerate microscopic states.
In terms of these states, the information loss paradox may be evaded by showing that a pure initial state collapses to a particular pure black hole microstate whose exact structure can be deduced by suitably subtle measurements.
Given the importance of the problem, it is crucial to ask:
What do pure microstates look like and what sorts of measurements can distinguish them from each other?
Here we report that in string theory almost no probes are able to differentiate the microstates of a black hole.
Thus, if spacetime is placed in a very heavy pure state, it will appear mixed --- {\em i.e.},\ like a black hole --- to almost all finite precision measurements.
This explains how the existence of pure underlying microstates and the absence of fundamental information loss are consistent with the semiclassical observation of a thermodynamic character to black holes.

We analyze so-called large black holes in AdS spacetimes.
These black holes have horizon size bigger than the scale set by the AdS curvature.
They are known to come into equilibrium with their thermal radiation because AdS geometries create an effective confining potential \cite{hawkingpage}.
Thus such black holes are stable, and we can ask how the underlying states that give rise to the large entropy can be identified via quantum mechanical probes.
We will examine black holes in five dimensions because it is known that string theory on $\AdS{5}\times S^5$ admits a dual description in terms of a superconformal $SU(N)$ gauge field theory propagating on a three-sphere \cite{ads}.
Here $N$ is related to the AdS length $\ell$ via the string coupling $g_s$ and the string length $\ell_s$ as $\ell^4 = 2\pi g_s N \ell_s^4$.
When the length scale $\ell$ is large, which we require, so too is $N$.

A large black hole in $\AdS{5}$ has mass $M \simeq r_0^2/G_5$, where the horizon scale $r_0$ is comparable to the $\AdS{}$ radius $\ell$ and $G_5 \simeq g_s^2 \ell_s^8 / \ell^5$ is the five-dimensional Newton constant.
The $\AdS{}$/CFT dictionary \cite{ads} maps an object with mass $M$ to an operator making states of energy $\Delta\simeq M\ell$ in the dual field theory.
Thus, a large black hole in $\AdS{5}$ corresponds to states with energy $\Delta\simeq N^2$.
Large black holes have an associated entropy $S = {\rm Area}/ 4G \simeq r_0^3/G_5 \simeq N^2$.
We therefore expect $e^{N^2}$ operators of conformal dimension (or energy) $N^2$ to describe the distinct pure microstates of the black hole.
Our purpose is to argue that there are almost no probes that can tell these microstates apart.   

What do the microstate operators look like?
Any gauge invariant operator of the dual field theory can be written as a Lorentz invariant polynomial in the elementary fields with traces imposing the gauge invariance.
Using just the scalar fields, one such operator might be
$$
\CO = (\tr{(D_\mu X) X^\dagger (D_\nu Y) Y^\dagger Z^\dagger}) (\tr{X^\dagger 
%Y^\dagger 
%Y^\dagger Y
 (D^\mu Z^\dagger) X (D_\rho X) 
 %Y
 }) (\tr{
 %Y^\dagger Y 
 %X 
 Y^\dagger Z Y (D^\nu D^\rho Z^\dagger) 
 %X^\dagger
 }) \cdots.
$$
Thus we can think of operators as long sentences built of words (traces), each of which is a string of letters in the alphabet provided by the elementary fields of the theory and their derivatives.
Black hole microstates are created by operators with very large dimension (energy) $\Delta \simeq N^2$ and hence involve polynomials of length of order $N^2$.
Theorems in information theory characterize the structure of such polynomials.
As $N$ becomes large, almost all operators belong to the ``typical set'' in which the string of letters is statistically random \cite{CT}.
Operators such as 
$$
\CO' = (\tr{XYXYXYXYXYXY
%XYXYXY
\ldots})(\tr{XYXYXYXYXYXY
%XYXYXY
\ldots})\ldots
$$
with discernible order are exponentially rare in the space of words.
Specifically, Sanov's theorem \cite{CT} shows that the probability of selecting an operator that deviates from a statistically random letter sequence is exponentially small in the dimension of the operators.
Thus almost all states with energy $N^2$ will look like statistically random strings with traces also distributed randomly.
Roughly, operators with $O(N)$ letters are interpreted as creating D-brane solitons, those with $O(\sqrt{N})$ letters create massive strings, and those with $O(1)$ letters create supergravity modes \cite{ads,states,corley}.
The operator microstate $\CO$ as a whole makes a state whose gravitational interpretation involves a conglomeration of all of these kinds of objects backreacting on spacetime.  

The above prescription does not specify an orthogonal basis of microstates.  In addition, supersymmetry is generically broken.  Thus there can be large mixings and renormalizations.
In addition, there is a constraint that a trace of more than $N$ $SU(N)$ matrices decomposes into a sum of smaller traces.
However, these effects will not change our main arguments because an operator constructed as a mixture of random polynomials will also look random.\footnote{It is amusing to note that operators that create the microstates of black holes have a dimension $\Delta \simeq N^2$ which is independent of the coupling. This suggests a mysterious absence of large renormalizations.}

A black hole microstate is created by applying the operators $\CO$ to the vacuum.
The state can be probed by computing correlation functions of field theory operators.
In spacetime, this corresponds to scattering probes off the pure quantum state that is supposed to have a macroscopic interpretation as a black hole.
Correlation functions are of the general form
\be
\bra{0} \CO_{\rm typical}^\dagger \CO_{\rm probe}^\dagger \CO_{\rm probe} \CO_{\rm typical} \ket{0}, \nn
\ee
where $ \CO_{\rm typical}$ is a typical (random polynomial) microstate operator, while $\CO_{\rm probe}$ is any probe state.
The probe operator will also be a polynomial in the elementary fields, which is interpreted as a gravity mode, string, D-brane, or black hole as described above.
Our main claim is that such correlators are almost independent of the precise microstate $\CO_{\rm typical}$.  To see this, first consider free field theory.
Then all contributions to the correlation function arise by Wick contracting fields with their conjugates --- {\em i.e.},\ free field propagators thread together the operators making the state and the probe.
Each pattern of Wick contractions makes a distinct contribution to the correlation function.
For clarity, consider small probes, {\em e.g.}, $\CO_{\rm probe} = \tr{X X}$.
Every contribution to the correlation function is determined entirely by the separation between the $X^\dagger$s in the state operator that contract with the probe, and by the sequence of letters separating the $X^\dagger$s.
Since the typical very large operator is a statistically random sequence of letters, every such pattern appears with an essentially universal frequency.
Thus the correlation function will be universal for all typical microstate operators.
Effectively, the correlation function is determined by the number of times sequences of letters in the probe operator appear as substrings of the typical operator.
Deviations from the universal result will therefore be exponentially small.

This argument extends to include interactions and to higher point correlators since the patterns of contractions between terms are still entirely controlled by the statistics of letters in large words.
In particular, it is evident that correlation functions involving small probes such as gravitons will simply be unable to piece together the details of a black hole microstate thus explaining how many pure states of quantum gravity can have the {\em same} long-wavelength description as a black hole.
As probe operators grow in energy, their ``resolving power'' increases, but it remains difficult to detect the black hole state.
For example, probes with energy $N$, roughly representing D-branes in spacetime, will have particularly large correlation functions if the probe operator appears identically as a term in the longer operator making up the microstate.
A D-brane probe can therefore separate the space of black hole microstates into classes that contain the same brane and those that don't.
Simple counting of states shows that almost all black hole states will not identically contain a particular brane, and thus the information obtained in this way will be limited.
More generally, a given microstate can be probed by another microstate --- {\em i.e.},\ a black hole can be probed by another black hole.
Since both microstates will be statistically random strings, this will generally give a universal result.
However, if the probe state is identical to the microstate, the response will be very large.
In this way, we see that in general it will take enormously many measurements to identify the actual microstate of a black hole, but it is significantly easier to identify what the state is not!
This is in precise analogy to probing the state of a gas of molecules.
It is easy to extract coarse-grained thermodynamic information (temperature, pressure, etc.), but to fix the state definitively requires an exact description of each particle in the gas.
In effect, because of their complexity, pure states are acting as mixed states.

The standard semiclassical notion of a horizon associated to a black hole at finite temperature is the spacetime analogue of the thermal-like coarse-grained description in field theory.
The properties of a microstate that we can readily measure in gauge theory correspond in the geometry to the global charges of a black hole. 
Our ideas are in beautiful accord with the ``fuzzball'' picture of black holes advocated in \cite{Mathur}.
Individual non-singular spacetime geometries without horizons are, in the conformal field theory, precisely the large operators that we have discussed.
The difficulty in telling two pure quantum states apart is related to the fact that a pair of given ``fuzzball'' geometries differ only at microscopic scales, and the measurement of different multipoles at infinity is not by itself sufficient to map out the total geography of the spacetime.
Interestingly, each microstate by itself can differ in detail from the exact black hole geometry out to the horizon scale, but in a way  that is not easily measurable by graviton probes.

To make the connection between gravity and gauge theory more explicit, we focus our attention on a particular subsector of operators in the dual field theory for which the geometries are known.
These are the operators that break half the supersymmetry (the so-called half-BPS operators).
The corresponding states are known to be stable and are not renormalized even in the presence of interactions.
Although these operators are atypical elements of the set of large operators, their properties are already meaningful for our purposes because they are on the verge of forming black holes \cite{myers}.
An illustrative class of such operators is the set of gauge-invariant polynomials constructed from a single scalar field, {\em e.g.},
\bea
\CO = \prod_{i=1}^k \tr{X^{n_i}}, && \sum_{i=1}^k n_i = \Delta, \nn
\eea
where $\Delta$ measures the dimension of the operator, or equivalently the energy carried by the corresponding state.
Generic half-BPS states of energy $\Delta$ are made from linear combinations of operators of this type.
By reasoning similar to that above, one can argue that the generic
operator of dimension $\Delta$ has a Boltzmann-like distribution of
trace lengths and coefficients.\footnote{A more efficient representation of half-BPS
states involves a map to a system of free fermions in an harmonic
potential, the excitations of which can be encoded in Young
diagrams \cite{corley,berenstein} and interpreted in terms of giant
gravitons, D3-branes wrapping three-spheres in $S^5$ or $AdS_5$
\cite{states,corley}. We will explain the structure describing
typical states in this basis in \cite{ourpaper}.}

\myfig{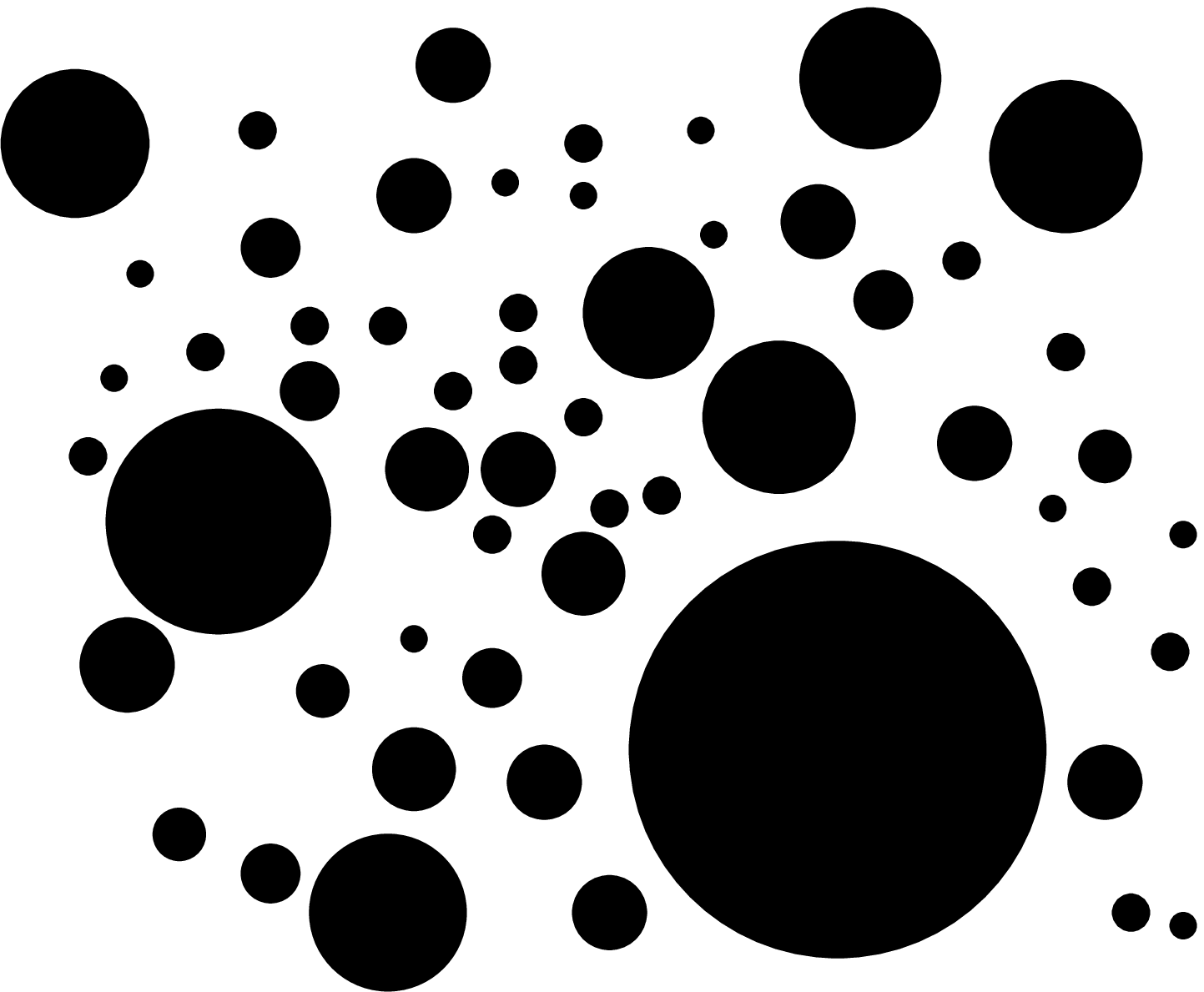}{6}{Foaming microstate geometry}

Using the techniques of \cite{llm} one can determine a precise spacetime geometry dual to each of these generic states.
All of them are asymptotically AdS and have no singularities or horizons.
However, near the origin of the space there is always a
two-dimensional plane with a complicated topological structure --- a
foam of spacetime ``bubbles'' --- related to the precise set of trace
lengths.\footnote{In the dual CFT, the corresponding two-plane arises 
from the phase space of a system of fermions in an harmonic potential 
\cite{corley,berenstein}. We will describe how the spacetime foam 
structure arises from a typical half-BPS state via its description in 
this fermion phase space in \cite{ourpaper}.}
In the complete ten-dimensional geometry relevant to string theory, these bubbles are different three-spheres expanding and shrinking at different points.   Pictorially, one may draw the diagram in Figure 1, where the black and white speckles indicate regions near the origin of the special plane where distinct  spheres in the geometries have collapsed.   Each typical half-BPS operator is dual to a geometry with a particular variegated pattern of speckles.  The relevant metrics can be written exactly, but we do not present them here due to their complexity.   One can show that the topological variations in the geometry occur at stringy and Planck scales (although they remain under some analytic control due to supersymmetry).
These exact geometries illustrate the point that we have made using dual gauge theory computations.
Identifying the exact patterns of speckles in Figure 1 requires enormously many, highly precise measurements, and graviton probes have wavelengths too long to resolve them.
To almost all probes Figure 1 will effectively look ``blurred'' --- {\em i.e.},\ the effective interaction of a probe with the spacetime could have been described equally well by an ensemble over geometries.   In this way, the effective thermodynamic character of gravity emerges.

\section*{Acknowledgements}
The material discussed in this essay has been presented at several
workshops, conferences and seminars including the Ohio Center for
Theoretical Physics Conference on Black Holes (V.B.),  the
String Cosmology workshop in Uppsala, Sweden (V.B.),  the Euclid `New 
Paths in Theoretical Physics' meeting in Trieste (V.J.),  the workshop on
`Gravitational Aspects of String Theory' at the Fields Institute (J.S.) 
and at the Institute for Studies in Theoretical Physics and
Mathematics in Tehran (V.J.) and the Perimeter Institute (J.S.). We
thank the organisers of these conferences, workshops and seminars for
the opportunity to present our work and the participants for
stimulating discussions. V.B. thanks the American University in 
Cairo for hospitality.
The current essay is a short version of a longer paper in
collaboration with Jan de Boer which will appear
elsewhere \cite{ourpaper} and which will contain a more complete list
of references on the subject. The work in Penn (V.B.\ and J.S.) is supported
by the DOE under grant DE-FG02-95ER40893, by the NSF under 
grant PHY-0331728 and OISE-0443607. The work of V.J.\ is supported by PPARC.

\end{document}